   \documentclass[aps,prb,twocolumn,groupedaddress,amsmath]{revtex4}
   \usepackage{bm}
\usepackage{graphicx,amssymb}
\usepackage{bm}
\usepackage{amsmath}
\usepackage{color}
\def\beq{\begin{equation}}
\def\eeq{\end{equation}}
\def\beqa{\begin{eqnarray}}
\def\eeqa{\end{eqnarray}}

\begin{document}
\draft
\title{Simple equation of state for hard disks on the hyperbolic plane}
\author{Mariano L\'opez de Haro}
\email{malopez@servidor.unam.mx}
\homepage{http://xml.cie.unam.mx/xml/tc/ft/mlh/} \affiliation{Centro
de Investigaci\'on en Energ\'{\i}a, Universidad Nacional Aut\'onoma
de M\'exico (U.N.A.M.), Temixco, Morelos 62580, M{e}xico}
\author{Andr\'es Santos}
\email{andres@unex.es} \homepage{http://www.unex.es/fisteor/andres/}
\author{Santos B. Yuste}
\email{santos@unex.es} \homepage{http://www.unex.es/fisteor/santos/}
\affiliation{Departamento de F\'{\i}sica, Universidad de
Extremadura, E-06071 Badajoz, Spain}
\date{\today}
\begin{abstract}
A simple equation of state for  hard disks on the hyperbolic plane
is proposed. It yields the exact second virial coefficient and
contains a pole at the highest possible packing. A comparison with
 another very recent theoretical proposal and simulation data is presented.
\end{abstract}
\maketitle

It is well known that
 hard-core systems\cite{Angel} represent
useful models that allow both the derivation of some rigorous
results in statistical mechanics as well as the computation of some
particular quantities such as virial coefficients. For monocomponent
hard-core  systems there cannot be a gas-liquid transition due to
the lack of an attractive part in the intermolecular potential, but
they show  crystalline and/or amorphous phases and the way these
phases arise and even their actual existence are still open
problems.

Insight into the thermodynamic behavior of hard-core systems has in
the past been sometimes gained by considering similar systems in
higher dimensions.\cite{highD} The rationale is that one may obtain
a rough idea of the thermodynamic behavior of say three-dimensional
hard-core fluids at high density by looking at models in higher
dimensions in which the same phenomenology is present but at a lower
density and thus the problem may become mathematically more
tractable. In a different context, but nevertheless dealing with a
somewhat related question, Modes and Kamien\cite{MK07} have
addressed the connection between the equation of state and
disordered packings of hard disks. Since in simulations of
monodisperse hard disks it is difficult to prevent crystallization,
the consideration of a curved surface, in this case a hyperbolic
plane with a curvature near a known regular tesselation, serves to
frustrate global crystalline order and allows them to find the
equation of state via molecular dynamics. Further, they have also
developed a free area theory for the packing derived from the nearby
tesselation. Studies in the hyperbolic plane in connection with
glass-forming liquids and bulk behavior of physical systems have
also recently been reported.\cite{Tarjus}

A few years ago we proposed a simple and accurate equation of state
for a hard-disk fluid.\cite{SHY95} This equation is built so as to
yield the exact second virial coefficient and also to have a single
pole singularity at the close-packing fraction, namely
\begin{equation}
\label{eq1}
{p}/{\rho k_BT}=\left[{1-b_2 \eta -  (1 -
b_2 \eta_{\text{max}})\eta^2/\eta_{\text{max}}^2}\right]^{-1},
\end{equation}
where  $p$ is the pressure, $\rho$ is the number density, $k_B$ is
the Boltzmann constant, $T$ is the absolute temperature, $b_2=2$ is
the reduced second virial coefficient, {$\eta=a_0(\sigma)\rho$} is
the packing fraction, with {$a_0(\sigma)=(\pi/4)\sigma^2$} the area
of a hard disk of diameter $\sigma$, and
$\eta_{\text{max}}=\sqrt{3}\pi/6\simeq 0.9069$ is the value
corresponding to crystalline close-packing. The major aim of this
Note is to provide an answer to the question of whether a proposal
such as Eq.\ (\ref{eq1}), properly generalized, may also be useful
for hard disks on the hyperbolic plane.

The development goes as follows.  First, we note that on a
two-dimensional manifold with constant intrinsic curvature $K<0$,
the area of a disk of diameter $\sigma$ is {$a_K(\sigma)=
2\pi|K|^{-1} \left[\cosh(|K|^{1/2}\sigma/2) - 1\right]$} and that
the packing fraction is  given by {$\eta=a_K(\sigma)\rho$}.
Furthermore, on such a manifold the associated (reduced) second
virial coefficient depends on {the (reduced) diameter
$|K|^{1/2}\sigma$, namely $b_2(|K|^{1/2}\sigma)=a_K(2
\sigma)/2a_K(\sigma)$}.\cite{MK07} Thus, for a given {value of the
reduced diameter $|K|^{1/2}\sigma$, the only other requirement in
our formulation is the corresponding value of $\eta_{\text{max}}$},
irrespective of the fact that the resulting configuration be an
ordered one or not. As far as we know, this value is only known
analytically for  {the so-called $\{p,q\}$ tesselations} (restricted
in the hyperbolic plane by the condition $1/p+1/q<1/2$), in which
the packing corresponding to the highest possible density and the
associated {reduced} diameter are given by
$\eta_{\text{max}}=\left[{\cos(\pi/q)}/{\sin(\pi/p)}-1\right]/({p/2-1-p/q})$
and
{$|K|^{1/2}\sigma=2\cosh^{-1}\left[{\cos(\pi/q)}/{\sin(\pi/p)}\right]$},
respectively.\cite{MK07}

\begin{figure}[h]
\includegraphics[width=\columnwidth]{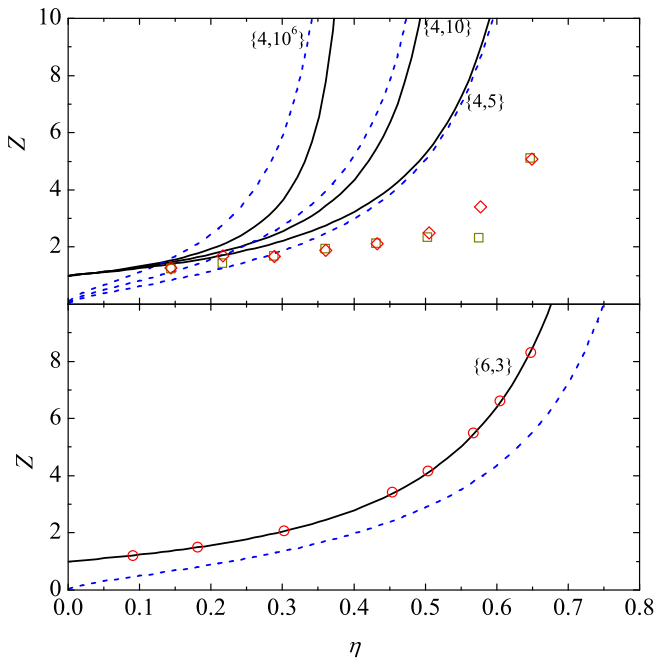}
\caption{{(Color online) Top panel: Compressibility factor $Z$ as a function of $\eta$ for the isostatic $\{4,10^6\}$, $\{4,10\}$, and $\{4,5\}$
tesselations ({$|K|^{1/2}\sigma=1.7628$}, $1.6169$, and $1.0613$, and $\eta_{\text{max}}=0.4142$, $0.5750$, and $0.7206$, respectively);
solid lines: Eq.\ (\protect\ref{eq1}); dashed lines: free area theory of Ref.\ \protect\onlinecite{MK07}; the symbols are simulation results
also obtained in Ref.\ \protect\onlinecite{MK07} for {$|K|^{1/2}\sigma=1.060$ (squares) and $|K|^{1/2}\sigma=1.062$} (diamonds).
Bottom panel:
Compressibility factor $Z$ as a function of $\eta$ for the (Euclidean) hexagonal $\{6,3\}$ tesselation
({$|K|^{1/2}\sigma=0$}, $\eta_{\text{max}}=\sqrt{3}\pi/6$); solid line: Eq.\ (\protect\ref{eq1}); dashed line: free area theory of Ref.\
\protect\onlinecite{MK07};
the symbols are the simulation results  from Ref.\ \protect\onlinecite{Erpenbeck}.} \label{fig1}}
\end{figure}

The top panel of Fig.\ \ref{fig1} shows the compressibility factor
$Z\equiv {p}/{\rho k_BT}$ as a function of the packing fraction
$\eta$, obtained from the free area theory of Modes and
Kamien\cite{MK07} and from the use of Eq.\ (\ref{eq1}), in the case
of three isostatic tesselations, that is, tesselations with $p=4$.
Clearly, the compressibility factors in both approaches are quite
close for high packing fractions, but they differ substantially at
low ones. We also compare the results for the packing fraction
dependence of $Z$ obtained with (to our knowledge) the only
presently available simulation data of Modes and Kamien\cite{note2}
for two values ({$|K|^{1/2}\sigma=1.062$ and
$|K|^{1/2}\sigma=1.060$}) of the hard-disk diameter which are close
to the one corresponding to the isostatic $\{4,5\}$ tesselation,
namely {$|K|^{1/2}\sigma=1.0613$}. The performance of both theories
with respect to the simulation is rather poor with, if any, a very
slight superiority of our approach. Nevertheless, no definite
statement about such a comparison should be made at this stage given
the very small number of particles (1 to 9) used in the simulations
and the intrinsic difficulties associated with simulations {on} the
hyperbolic plane.\cite{Tarjus}

To further test the performance of both theories in a more
controllable situation, in the bottom panel of Fig.\ \ref{fig1} we
present the results obtained in the Euclidean limit
({$|K|^{1/2}\sigma=0$}) using the hexagonal $\{6,3\}$ tesselation.
Of course in this limit in our approach we recover the results of
our earlier proposal for the equation of state of a hard-disk
fluid,\cite{SHY95} which we had proved to be rather accurate. The
free area theory,\cite{MK07} on the other hand, performs very
poorly.

While the philosophy behind the derivation of our equation of state
is totally different from that of the free area theory of Modes and
Kamien,\cite{MK07} they both share the property of having a pole at
$\eta=\eta_{\text{max}}$. In fact, from Eq.\ (\ref{eq1}) it follows
that {$Z\simeq [2- b_2(|K|^{1/2}\sigma) \eta_{\text{max}}]^{-1}
(1-\eta/\eta_{\text{max}})^{-1}$} while the free area theory yields
$Z_{\text{free area}} \simeq 2 (1-\eta/\eta_{\text{max}})^{-1}$ for
$\eta$ close to $\eta_{\text{max}}$. Note that if
{$|K|^{1/2}\sigma=1.0613$ then $[2- b_2(|K|^{1/2}\sigma)
\eta_m]^{-1}\simeq 2.20$}. This explains why at high packing
fractions the numerical values of the compressibility factor
obtained with both approaches are not all that different. One should
point out that the free area theory has been constructed with the
particular aim of performing well at high densities while ours
attempts to capture both the low and high density limits. On the
other hand,  $Z_{\text{free area}}$ is nonanalytic at $\eta=0$ (for
the above {value of $|K|^{1/2}\sigma$}, $Z_{\text{free area}}=1.262
\eta^{1/2}+1.4328 \eta + 1.5941 \eta^{3/2}+ \cdots$ ) and
incorrectly predicts $Z_{\text{free area}}(0)=0$, whereas by
construction $Z$ given by Eq.\ (\ref{eq1}) yields the exact second
virial coefficient {$b_2(|K|^{1/2}\sigma)$} and also allows us to
estimate the higher order virial coefficients
{$b_n(|K|^{1/2}\sigma)$}. Thus, from Eq.\ (\ref{eq1}) and again for
{$|K|^{1/2}\sigma=1.0613$}, we get $b_3(1.0613)=3.55$,
$b_4(1.0613)=5.36$, and $b_5(1.0613)=7.76$. Comparing with the
numerical values obtained by Modes and Kamien\cite{note2} for
{$|K|^{1/2}\sigma=1.10$}, namely $b_3(1.10)=3.39$, $b_4(1.10)=4.62$,
and $b_5(1.10)=5.83$, we find that our estimates are of the right
order of magnitude although their accuracy seems to be worse than
the one we got for the Euclidean case.\cite{SHY95} As a final
comment one should add that Eq.\ (\ref{eq1}) {may lead to
(unphysical) negative} values of $Z$ in the interval
{$\eta_{\text{max}}/\left[b_2(|K|^{1/2}\sigma)\eta_{\text{max}}-1\right]<\eta<\eta_{\text{max}}$}
if {$b_2(|K|^{1/2}\sigma)\eta_{\text{max}}>2$}. This may well be a
limitation of our formulation but the point certainly deserves
further study.

In summary, in this Note we have presented an extension of our
former equation of state for a hard-disk fluid to deal with the same
system {on} the hyperbolic plane. In contrast with the free area
theory of Ref.\ \onlinecite{MK07}, which is rather more complicated
and devised to perform well in the high density region only, the
main assets of our proposal are its simplicity and the fact that it
caters both for low and high packing fractions. Our results indicate
that it might be as accurate and even improve on the performance of
the free area theory of Modes and Kamien.\cite{MK07} A deeper
assessment of its full value is not possible at this stage due to
the lack of simulation results for different values of
{$|K|^{1/2}\sigma$}. We hope that our simple theory will encourage
the performance of such simulations.

We want to thank C. D. Modes and R. D. Kamien for kindly providing
us with their simulation data and their values for the virial
coefficients. The work of A.S. and S.B.Y. has been supported by the
Ministerio de Educaci\'on y Ciencia (Spain) through Grant No.
FIS2007-60977 (partially financed by FEDER funds) and by the Junta
de Extremadura through Grant No. GRU08069. M.L.H. acknowledges the
partial financial support of DGAPA-UNAM under project IN-109408.


\begin{references}
\bibitem{Angel}For a very recent review of the properties of these systems one may look at \textit{Theory and Simulation
of Hard-Sphere Fluids and Related Systems}, edited by A. Mulero
(Springer, Berlin, 2008).
\bibitem{highD}
A few illustrative papers are H. L. Frisch and J. K. Percus, Phys.
Rev. A \textbf{35}, 4696 (1987);  M. Baus and J. L. Colot,
\textit{ibid.} \textbf{36}, 3912 (1987); H. L. Frisch and J. K.
Percus, Phys. Rev. E \textbf{60}, 2942 (1999); G. Parisi and F.
Slanina, \textit{ibid.} \textbf{62}, 6554 (2000); A. Santos  and M.
L\'opez de Haro, \textit{ibid.} \textbf{72}, 010501(R) (2005); F.
Zamponi, Phil. Mag. \textbf{87}, 485 (2007); and R. Rohrmann, M.
Robles, M. L\'opez de Haro, and A. Santos, J. Chem. Phys.
\textbf{129}, 014510 (2008), where many more references to work on
hard-core systems in $D$ dimensions may be found.

\bibitem{MK07}
C. D. Modes and R. D. Kamien, Phys. Rev. Lett. \textbf{99}, 235701
(2007); Phys. Rev. E. \textbf{77}, 041125 (2008).

\bibitem{Tarjus} F.
Sausset and G. Tarjus, J. Phys. A: Math. Gen. \textbf{40}, 12873
(2007); preprint arXiv:0805.1475; F. Sausset, G. Tarjus, and P.
Viot, preprint arXiv:0805.2819.

\bibitem{SHY95}
A. Santos, M. L\'opez de Haro, and S. B. Yuste, J. Chem. Phys.
\textbf{103}, 4622 (1995);
 M. L\'opez de Haro, A. Santos,  and S. B. Yuste,  Eur. J. Phys. \textbf{19}, 281
 (1998).


\bibitem{note2}
C. Modes and R. D. Kamien, private communication.


\bibitem{Erpenbeck}
J. J. Erpenbeck and M. Luban, Phys. Rev. A. \textbf{32}, 2920 (1985).

\end{references}
\end{document}